\documentclass[twocolumn,showpacs,preprintnumbers,amsmath,amssymb,pre]{revtex4}
\usepackage{graphicx}
\usepackage{dcolumn}
\usepackage{bm}

\newcommand{\nil}{\hspace*{0em}}
\begin{document}
\draft
\title{Orbital order, anisotropic spin
couplings, and the spin-wave spectrum of the ferromagnetic Mott
insulator YTiO$_{\text{3}}$}
\author{Robert Schmitz,$^{1,2}$ Ora Entin-Wohlman,$^2$
Amnon Aharony,$^2$ and Erwin M\"uller-Hartmann$^1$}
\affiliation{$^1$Institut f\"ur Theoretische Physik, Universit\"at
zu K\"oln, Z\"ulpicher Stra{\ss}e 77, 50937 K\"oln,
Germany\\$^2$School of Physics and Astronomy, Raymond and Beverly
Sackler Faculty of Exact Sciences, Tel Aviv University, Tel Aviv
69978, Israel}

\date{\today}

\begin{abstract}
Using a point-charge calculation of the electrostatic crystal
field, we determine the non-degenerate orbital ground state of the
ferromagnetic Mott insulator YTiO$_{\text{3}}$, which is found to
agree perfectly with experiment. Based on the orbital order, we
obtain by perturbation theory an effective spin Hamiltonian that
describes the  magnetic superexchange between nearest-neighbor Ti
ions. The superexchange Hamiltonian includes, in addition to the
isotropic Heisenberg coupling, antisymmetric
(Dzyaloshinskii-Moriya) and symmetric anisotropy terms,  caused by
the spin-orbit interaction on the Ti ions. We find ferromagnetic
Heisenberg couplings for Ti--Ti bonds in the crystallographic $ab$
planes, but antiferromagnetic ones for Ti--Ti bonds between
planes, in contradiction with experiment (which gives
ferromagnetic couplings for both). Difficulties in calculating
realistic values for the isotropic couplings of YTiO$_3$ have been
already reported in the literature. We discuss possible origins
for these discrepancies. However, the much smaller values we
obtain for the symmetric and antisymmetric anisotropies may be
expected to be reliable. We therefore combine the
experimentally-deduced isotropic coupling with the calculated
anisotropic ones to determine the magnetic order of the Ti ions,
which is found to be in satisfactory agreement with experiment.
Based on this magnetic order, we  derive the spin-wave spectrum.
We find an acoustic branch with a very small zone-center gap and
three optical spin-wave modes with sizeable zone-center gaps. The
acoustic branch reproduces the one reported in experiment, and the
optical ones are in a satisfactory agreement with experiment, upon
a proper folding of the magnetic Brillouin zone.
\end{abstract}


\maketitle

\section{Introduction}

The perovskite Ti oxides have attracted much interest since these
strongly-correlated electronic systems  possess orbital and
magnetic degrees of freedom which are coupled together (for a
review, see Ref.~\cite{mochi1}). A prominent member of this family
is the {\em ferromagnetic} Mott insulator YTiO$_3$. The Curie
temperature of this compound is $T_{\text{C}}=30$ K and the
ordered ferromagnetic moment, which is oriented along the
crystallographic $c$ axis, is 0.84 $\mu_{\text{B}}$ \cite{garret}.
Another experiment \cite{ulrich} reported $T_{\text{C}}=27$ K. Due
to spin canting, there are also a small G-type antiferromagnetic
moment along the $a$ axis and a small A-type one along the $b$
axis, which at $T=10$ K amount to 0.08 $\mu_{\text{B}}$ and 0.05
$\mu_{\text{B}}$, respectively  (the $c$-axis ferromagnetic moment
being 0.54 $\mu_{\text{B}}$ at that temperature, which is
extrapolated to 0.72 $\mu_{\text{B}}$ at zero temperature)
\cite{ulrich}.

Several previous calculations aiming to explain  YTiO$_3$ and the
doped series La$_{1-x}$Y$_x$TiO$_3$, respectively, have failed to
achieve a consistent description of the experimentally observed
orbital and magnetic ordering
\cite{ulrich,mochi2,mochi3,solovyev}. A recent GGA+U (generalized
gradient approximation + local Coulomb repulsion) study
\cite{okatov} has produced  the correct orbital and magnetic
ground state of YTiO$_3$, but has not provided quantitative
estimates for the superexchange couplings between nearest-neighbor
Ti ions. These couplings are required in order to understand
quantitatively the magnetic structure and the spin-wave spectrum
observed in experiment \cite{ulrich}.

From the microscopic point of view,  YTiO$_3$ is quite similar to
the {\em antiferromagnetic} Mott insulator LaTiO$_3$:  In both
compounds there is a single electron in the 3$d$ shell of Ti,  and
both have the same space group,  $Pbnm$. It is worth noting in
this connection that the magnetic structure of  LaTiO$_3$ is also
quite complicated: Experiment has indicated  a predominant
antiferromagnetic G-type moment along the $a$ axis and a small
ferromagnetic moment along the $c$ axis \cite{cwik}, and a recent
theory \cite{us} has predicted  a small A-type moment along the
$b$ axis.  We have recently presented a detailed model for
LaTiO$_3$ \cite{us}, which proved successful in describing the
orbital and magnetic ordering of that material and provided the
superexchange couplings and the spin-wave dispersion measured in
experiment \cite{keimer}. Because of the apparent similarity
between YTiO$_3$ and LaTiO$_3$, one may hope that the same model
will explain the former as well. In this paper we carry out such
investigation. Unfortunately, our model does not yield the correct
isotropic Heisenberg superexchange coupling between
nearest-neighbor Ti ions. The reason is  that there are both
ferromagnetic and antiferromagnetic contributions to that
coupling, which have roughly the same order of magnitude. Our
approximations cannot resolve the competition between them to a
sufficient precision. A similar problem has been reported in
Ref.~\cite{ulrich}.

Superexchange couplings are customarily derived perturbatively,
assuming that  the hopping matrix elements are smaller than the
on-site excitation energies. It has been found in other systems
\cite{yild} that such calculations yield inaccurate values for the
leading Heisenberg (isotropic) couplings, but are quite reliable
for the much smaller anisotropic ones. It seems therefore
reasonable to combine  the experimental information on the
isotropic couplings together with the calculated values of the
anisotropic ones, in order to determine the magnetic structure of
the ground state and the spin excitations. This procedure will be
adopted in this paper.

We begin in Sec.~\ref{model} with a  point-charge calculation of
the electrostatic crystal field due to all ions of the solid. The
ground state of this crystal field determines the orbital order of
the Ti ions. This orbital ordering agrees extremely well with the
one detected experimentally  \cite{itoh,aki}. We then use a
Slater-Koster parametrization to compute the effective Ti--Ti
hopping matrix elements. The Coulomb correlations on the doubly
occupied $d$ shells are fully taken into account in terms of
Slater integrals. Having thus obtained an effective microscopic
Hamiltonian for the Ti ions, we derive in Sec. ~\ref{secouplings}
the superexchange spin couplings between nearest-neighbor  Ti
ions, employing perturbation theory to leading (second) order in
the hopping matrix elements, and up to second order in the
spin-orbit interaction on the Ti's. In this way we obtain, beside
the isotropic superexchange coupling alluded to above, the
antisymmetric (Dzyaloshinskii-Moriya) and symmetric superexchange
anisotropies. Replacing the isotropic coupling by the
experimentally-determined one, we calculate in Sec.~\ref{magngs}
the magnetic order of the classical ground state of the Ti ions
and in Sec.~\ref{spinwaves} the spin-wave spectrum. The magnetic
structure of the classical ground state is shown to be in
satisfactory agreement with experiment \cite{ulrich}. The
spin-wave calculation reproduces an acoustic branch of the
spin-wave dispersion which has been detected by neutron scattering
\cite{ulrich}. This branch has a very small zone-center gap, and
is almost isotropic in the magnetic Brillouin zone. In addition,
we identify three optical spin-wave branches with considerable
zone-center gaps. The experimental dispersion has been plotted as
a single branch over the magnetic Brillouin zone (MBZ) of a pure
ferromagnet \cite{ulrich}. However, YTiO$_3$ is a {\em{canted}}
ferromagnet, for which the MBZ is four times smaller. We therefore
prefer to re-plot the experimental spin-wave data according to the
actual MBZ, i.~e., to fold back the experimental data from the MBZ
of the purely ferromagnetic case. When this procedure is adopted,
one obtains a satisfactory agreement between the optical branches
and experiment. In Sec. \ref{summary} we summarize our results and
compare our picture of YTiO$_3$ with the ones given previously in
the literature \cite{ulrich,mochi2,mochi3,solovyev}.

\section{The model} \label{model}

\subsection{The crystal field}

\begin{figure}
\includegraphics[width=7cm]{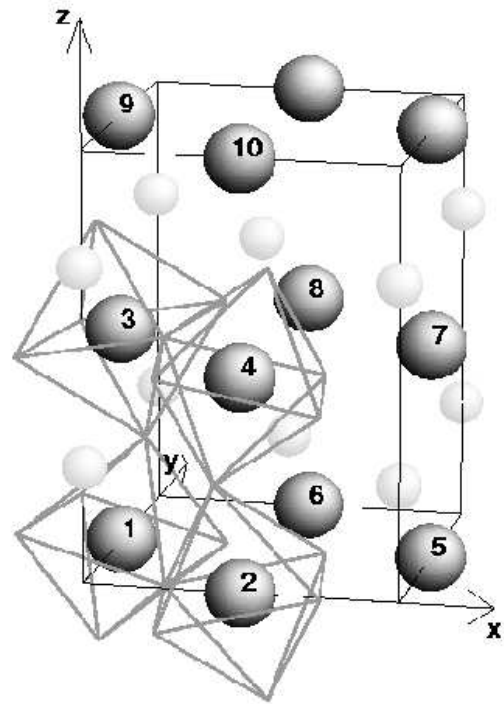}
 \caption{The crystallographic structure of YTiO$_3$. The ten Ti
ions, which constitute the twelve inequivalent nearest-neighbor
Ti--Ti bonds are enumerated. For simplicity, oxygen octahedra are
only shown around four Ti sites. Y ions from two layers are shown
as small spheres. }\label{fig1}
\end{figure}

As is mentioned above, there is a single electron in the 3$d$
shell of the Ti ions in the ground state (YTiO$_3$ has  the
valences Y$^{3+}$Ti$^{3+}$(O$^{2-}$)$_3$). The unit cell, shown in
Fig.~\ref{fig1}, contains four Ti ions and twelve inequivalent
nearest-neighbor Ti--Ti bonds. The crystal has the symmetry of the
space group $Pbnm$ (No. 62 in Ref.~\cite{hahn}). The structural
data (taken at $T=2\,$K) are given in Table \ref{tab1}
\cite{komarek}. In order to use the symmetries of the space group,
it is convenient to employ the orthorhombic orthonormal basis for
the Ti-$d$ orbitals,
\begin{equation}
\big|xy\big>,\big|2z^2\big>,\big|yz\big>,\big|xz\big>,\big|x^2-y^2\big>,
\label{dbasis}
\end{equation}
where the $x$, $y$ and $z$ axes correspond to the crystallographic
$a$, $b$ and $c$ axes.

\setlength{\tabcolsep}{0.05cm}
\renewcommand{\arraystretch}{1.3}
\begin{table}
\caption{The structural parameters of YTiO$_3$ at $T=2$ K
\cite{komarek}.}
\begin{ruledtabular}
\begin{tabular}{ll|ll}
$a$ & 5.32260 \AA & $x_{\text{O1}}$ & 0.12133 \\
$b$ & 5.69517 \AA & $y_{\text{O1}}$ & 0.45702 \\
$c$ & 7.59622 \AA & $x_{\text{O2}}$ & 0.69010 \\
$x_{\text{Y}}$ & 0.97762 & $y_{\text{O2}}$ & 0.30919 \\
$y_{\text{Y}}$ & 0.07398 & $z_{\text{O2}}$ & 0.05770
\end{tabular}
\end{ruledtabular}
\label{tab1}
\end{table}

\begin{figure}
\includegraphics[width=7cm]{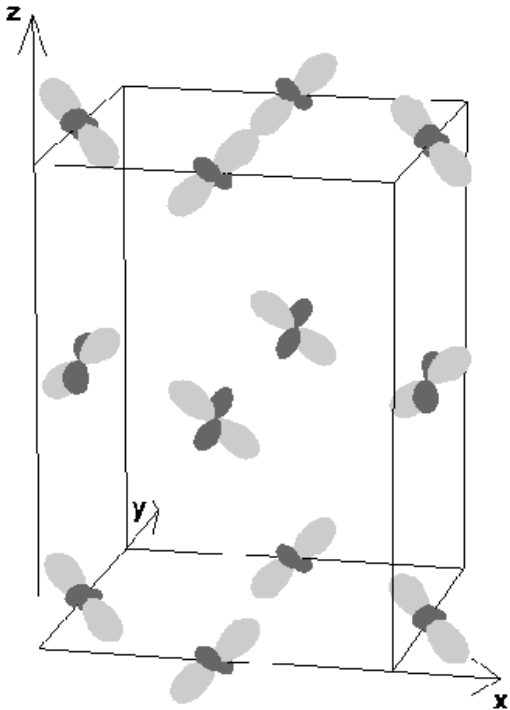}
\caption{The orbital order of the Ti ions in YTiO$_3$.}
\label{fig2}
\end{figure}

Using the structural data listed in Table \ref{tab1}, we have
calculated the spectrum and the eigenstates of Ti ion No.~1 (see
Fig.~\ref{fig1}), employing a point-charge calculation of the
static crystal-field Hamiltonian. This calculation uses the full
Madelung sum over the crystal, which is evaluated as an Ewald sum
\cite{ewald}. It requires the second moment, $\big<r^2\big>$, and
the fourth moment, $\big<r^4\big>$, of the effective ionic radius
of the Ti$^{3+}$-ion. We have used the values $\big<r^2\big>=0.530
\;\text{\AA}\nil^2$ and $\big<r^4\big>=0.554 \; \text{\AA}\nil^4$
\cite{altsh}. The results of the crystal-field calculation, which
are listed in Table \ref{tab2}, exhibit a $t_{2g}$ splitting
scheme where a non-degenerate ground state is clearly separated
from the excited states. This ground state orbital, which gives
rise to orbital ordering, is given by the first line in Table
\ref{tab2} and depicted in Fig.~\ref{fig2}. It agrees very well
with data obtained from nuclear magnetic resonance (NMR)  and
polarized neutron diffraction experiments \cite{itoh,aki}.

\begin{table}
\caption{The static crystal field for Ti$^{3+}$ (site 1): Spectrum
and eigenstates in the orthorhombic basis for the $d$ basis, see
Eq. (\ref{dbasis}).} \label{tab2}
\begin{ruledtabular}
\begin{tabular}{r|rrrrr}
--0.458 eV & (--0.181,& 0.295,& 0.488,& --0.542,& 0.590)\\
\hline --0.308 eV & (--0.081,& --0.412,& 0.529,& 0.653,& 0.343) \\
\hline --0.181 eV & (\hspace*{0.5em}0.444,& 0.266,& 0.654,&
--0.017,& --0.552) \\ \hline
 0.407 eV & (\hspace*{0.5em}0.761,& 0.302, &--0.231, &0.222,& 0.477) \\ \hline
 0.540 eV & (--0.430,& 0.762,& --0.039,& 0.480,& --0.040)
\end{tabular}
\end{ruledtabular}
\end{table}

\subsection{The Hamiltonian}

We next construct the microscopic Hamiltonian pertaining to the Ti
ions, from which we  obtain perturbatively the superexchange spin
couplings. The calculation is carried out for a two-site cluster,
consisting of two nearest-neighbor Ti ions, denoted by $m$ and
$n$.

The unperturbed part of the Hamiltonian contains the static
crystal field, $H_{mn}^{\text{cf}}$, and the intra-ionic Coulomb
correlations of a doubly occupied $d$ shell, $H_{mn}^{\text{c}}$,
\begin{equation}
H_{mn}^0=H_{mn}^{\text{cf}}+H_{mn}^{\text{c}}. \label{unperturbed}
\end{equation}
The perturbation calculation carried out below involves two
sectors, which together span the Hilbert space of the cluster: A
Ti$^{3+}$ sector, in which both Ti ions are trivalent, and a
Ti$^{2+}$ sector, where one of the Ti ions is divalent (two $d$
electrons on the same site) and the other is tetravalent (an empty
$d$ shell). The ground state of $H_{mn}^0$ belongs to the
Ti$^{3+}$ sector, where both Ti ions are in the one-particle
ground state of $H_{mn}^{\text{cf}}$ (modulo spin up or down on
each site), leading to a four-fold degeneracy of the ground state
of the cluster. The spectrum of $H_{mn}^0$ is found by applying
$H_{mn}^{\text{cf}}$ on the Ti$^{3+}$ sector, and both
$H_{mn}^{\text{cf}}$ and $H_{mn}^{\text{c}}$ on the Ti$^{2+}$
sector. $H_{mn}^{\text{c}}$ is parametrized in terms of the Slater
integrals $F^2$ and $F^4$ \cite{slater}, and the effective Ti--Ti
charge-transfer energy $U_{\text{eff}}$. This energy is the
difference between the four-fold degenerate ground state of the
cluster (which is the lowest level of the Ti$^{3+}$ sector) and
the lowest level of the Ti$^{2+}$ sector (where
$H_{mn}^{\text{cf}}$ and $H_{mn}^{\text{c}}$ are diagonalized
simultaneously). We use $F^2=8 F^4/5=8.3\,\text{eV}$ from an
atomic Hartree-Fock calculation \cite{watson}, and
$U_{\text{eff}}=3.5\,\text{eV}$ from the analysis of the
photoemission spectra and first-principles calculations
\cite{bocquet}. We note that the charge-transfer energy
$U_{\text{eff}}$ might have a considerable uncertainty, and in
particular may be lower than 3.5 eV. For example, there is a
strong resonance in Raman spectroscopy \cite{ulrich2} at a laser
frequency of 2.54 eV. Since experiment indicates that the
resonance is mainly caused by processes involving two Ti sites, it
may well be that it yields a lower value for $U_{\text{eff}}$.
Further first-principles calculations and a comparison with
optical-conductivity data are required in order to determine more
precisely $U_{\text{eff}}$.

The perturbation part of the Hamiltonian, $V_{mn}^{\nil}$,
consists of an effective Ti--Ti tunnelling term,
$H_{mn}^{\text{tun}}$, and the on-site spin-orbit interaction,
$H_{mn}^{\text{so}}$,
\begin{equation}
V_{mn}^{\nil}=H_{mn}^{\text{tun}}+H_{mn}^{\text{so}}.\label{perturbed}
\end{equation}
The tunnelling Hamiltonian is given in terms of an effective
hopping matrix, $t_{mn}$, between the $m$ and the $n$ Ti ions,
\begin{equation}
H_{mn}^{\text{tun}}=\sum_{ij}\sum_\sigma t_{mn}^{ij}
d_{mi\sigma}^{\dag}d_{nj\sigma}^{\hspace*{0em}} +\text{H.\,c.},
\label{Htun}
\end{equation}
where $d_{mi\sigma}^{\dagger}$ ( $ d_{mi\sigma}^{\nil})$ creates
(destroys) an electron with spin $\sigma$ in the $i$-th
eigen-orbital of $H^{\rm cf}_{mn}$ at site number $m$ (see Table
\ref{tab2}). The spin-orbit coupling is given by
\begin{equation}
H_{mn}^{\text{so}}=\lambda \hspace*{-0.5em} \sum_{k=m,n}
\hspace*{-0.5em} {\mathbf{l}}_k \cdot {\mathbf{s}}_k,
\end{equation}
where ${\mathbf{l}}_k$ denotes the angular momentum operator of
the Ti ion at the $k$ site, ${\mathbf{s}}_k$ is its spin operator,
and $\lambda $  is the spin-orbit coupling strength. We use
$\lambda=18\,\text{meV}$ \cite{imada3}.

The dominant hopping process between two nearest-neighbor Ti ions
is mediated via the oxygen ion which is nearest to both of them.
Let $t^{i\alpha}_m $ be the hopping matrix element of an electron
in the $p$ orbital $\alpha$ on the oxygen ion  into the $i$ state
of the  Ti ion located at $m$. The effective hopping between the
Ti ions is then given by
\begin{equation}
 t_{mn}^{ij}=-\frac{1}{\Delta_{\text{eff}}}
 \sum_{\alpha} t^{i\alpha}_m
 t^{j\alpha}_n=t_{nm}^{ji}.\label{EFFTUN}
\end{equation}
Here, $\Delta_{\text{eff}}$ is the charge-transfer energy, which
is required to put an electron from an O ion on a Ti ion, and
$\alpha$ denotes one of the three $p$ orbitals on the oxygen (in
orthorhombic coordinates),
\begin{equation}
\big|x\big>,\big|y\big>,\big|z\big>. \label{pstates}
\end{equation}
(Modifications of this basis due to the crystal field are ignored,
since the crystal field splitting is expected to be small compared
to the Ti-O charge-transfer energy.)

  Using the structural data from Table \ref{tab1},
together with elementary geometric considerations, the Ti--O
hopping matrix elements can be expressed in terms of the
Slater-Koster parameters $V_{pd\sigma}$ and $V_{pd\pi}$
\cite{harrison}. We use the values $V_{pd\sigma}=-2.3$ eV,
$V_{pd\pi}=1.1$ eV, and $\Delta_{\text{eff}}=5.5$ eV
\cite{bocquet}, in conjunction with Eq.~(\ref{EFFTUN}) to compute
the effective hopping matrices pertaining to the unit cell. The
results are listed in Table \ref{tab3}, which also gives the
symmetry properties of the hopping matrices between different
Ti--Ti bonds. The four Ti sites of the unit cell form twelve
nearest-neighbor Ti--Ti bonds which are inequivalent, i.\,e., they
do not evolve from each other by Bravais translations. These bonds
connect the ten Ti ions indicated in Fig.~\ref{fig1}. By the
symmetry operations of the space group $Pbnm$, the eight effective
hopping matrices between Ti ions belonging to the same $ab$ plane
and the four matrices for Ti--Ti bonds along the $c$ direction,
respectively, can be expressed by a single matrix each. For
example, all twelve hopping matrices are given by the two matrices
for the Ti--Ti bonds $mn=12$ (planar) and $mn=13$ (inter-planar),
respectively.

\begin{table}
\caption{The effective Ti--Ti hopping matrices for the $d$
eigen-orbitals of the crystal field from Table \ref{tab2}; values
are given in eV. The rows and the columns are ordered beginning
with the ground state of the crystal field (index 0), continuing
with the first excited state (index 1), etc. The matrix $t_{13}$
is symmetric because of a mirror plane, see Ref.~\cite{us}.}
\label{tab3}
\begin{ruledtabular}
\begin{tabular}{l}
Planar \\ \hline
$t_{12}^{\nil}=t_{16}^{t}=t_{25}^{\nil}=t_{65}^{t}=t_{34}^{\nil}
=t_{38}^{t}=t_{47}^{\nil}=t_{87}^{t}$\\[1ex]
$=\left[
\begin{array}{rrrrr}
-0.062& -0.206& 0.033& -0.026& -0.012\\
0.007& -0.015& 0.006& -0.086& 0.114\\
0.130& -0.077& -0.125& 0.149& -0.203\\
-0.202& 0.030& 0.092& 0.453& -0.632\\
-0.036& 0.008& 0.024& 0.031& -0.044
\end{array}
\right]_{\vspace*{1mm}}$ \\ \hline Inter-planar \\ \hline
$t_{13}^{\nil}=t_{24}^{\nil}=t_{39}^{\nil}=t_{410}^{\nil}$\\[1ex]
$=\left[
\begin{array}{rrrrr}
0.086& -0.009& 0.101& -0.024& -0.085\\
-0.009& 0.160& 0.043& 0.126& 0.227\\
0.101& 0.043& 0.119& -0.048& -0.159\\
-0.024& 0.126& -0.048& -0.107& -0.263\\
-0.085& 0.227& -0.159& -0.263& -0.607
\end{array}
\right]_{\vspace*{1mm}}$
\end{tabular}
\end{ruledtabular}
\end{table}

\subsection{The Ti--O hybridization}

Our model does not include  the covalent contribution to the
crystal field, arising from hybridization between the Ti--3$d$ and
O--2$p$ states. This mechanism mixes excited states of the static
crystal-field into the Ti$^{3+}$ ground state, i.\,e., there is an
admixture of Ti$^{2+}$ states accompanied by an admixture of holes
on the oxygen sites.

Following Refs.~\cite{bocquet} and \cite{arima}, we may estimate
the effect of the $pd$ hybridization. When that hybridization is
absent, the effective parameter $U_{\text{eff}}$ defines the
energy difference between the ground state of the Ti$^{3+}$ sector
and the lowest state of the Ti$^{2+}$ sector in a two-site cluster
consisting of two Ti ions. When the $pd$ hybridization is present,
these two types of $d$ states correspond to two bands, from which
two $pd$ hybridized bands evolve according to the covalent crystal
field. These hybridized bands have, in general,  significant
dispersion: Their peak-to-peak separation is given by the band gap
$E_{\text{gap}}$=1.8 eV \cite{bocquet}, and the distance between
the band edges is given by the optical gap $E_{\text{opt}}$=1.0 eV
\cite{arima}, which is experimentally observed as the Mott gap.
The mean bandwidth between the two $pd$ hybridized bands is then
$W=E_{\text{gap}}-E_{\text{opt}}=0.8$ eV. These bands are not as
dispersive as in the case of LaTiO$_3$, where the mean bandwidth
is $W=1.4$ eV \cite{saitoh}.

Nevertheless,  given this  dispersion of the bands one may wonder
whether a localized picture is appropriate, even approximately,
for the YTiO$_{3}$ system. In order to study this point, we have
analyzed the covalent crystal field of a cluster consisting of a
single Ti ion, and the six oxygen ions predominantly hybridized
with it (the calculation has been carried out for Ti number 1 in
Fig.~\ref{fig1}). This is accomplished by diagonalizing  the
Hamiltonian
\begin{eqnarray}
H_{pd}^{\nil}=
H_{\nil}^{\text{cf}}+H_{\nil}^{\text{c}}+H_{pd}^{\text{tun}},\label{Hpd}
\end{eqnarray}
for a TiO$_{6}$--cluster. Here $H^{\text{cf}}$ describes the
static crystal field, $H^{\text{c}}$ is the Coulomb interaction,
and $H_{pd}^{\text{tun}}$ is the $pd$--tunnelling,
\begin{equation}
H_{pd}^{\text{tun}}=\sum_{ni\alpha \sigma}
\tilde{t}_{1n}^{i\alpha} d_{1i\sigma}^{\dag}p_{n\alpha
\sigma}^{\nil} +\text{H.\,c.,}\label{pdtun}
\end{equation}
where $p_{n\alpha \sigma}$ destroys an electron on the $n$-th
oxygen site with spin $\sigma$ in the $\alpha$--orbital, given in
Eq.~(\ref{pstates}). As in the calculation of the Ti--Ti hopping
amplitudes, the $pd$ hopping amplitudes,
$\tilde{t}_{1n}^{i\alpha}$, are expressed in terms of the
Slater-Koster parameters $V_{pd\sigma}$ and $V_{pd\pi}$, using the
structural data of Ref.~\cite{komarek}.

The entire space of the basis states of the TiO$_{6}$--cluster
consists of a Ti$^{3+}$ sector where the $p$ orbitals are all
occupied, and a Ti$^{2+}$ sector where there is a hole in one of
the $p$ orbitals. The eigenstates of the Hamiltonian (\ref{Hpd})
have the form
\begin{equation}
\big|\psi\big>=\sqrt{2-n_d}\,\big|d^1\big>+\sqrt{n_d-1}\,\big|d^2\big>,
\label{clusterlc}
\end{equation}
where $n_d$ is the occupation number of the Ti-$d$ shell ($1 \leq
n_d \leq 2$), $\big|d^1\big>$ is a state with a single electron in
the $d$ shell and fully occupied $p$ shells on the surrounding
oxygen ions, and $\big|d^2\big>$ is a state with two electrons in
the $d$ shell and a hole in the $p$ shell of one of the oxygen
ions. We find that in the ground state $n_d=1.330$ , i.\,e., there
is a $p$ hole on one of the neighboring oxygens with probability
of $33.0$ \%.

This calculation allows for the analysis of the eigenstates of the
combined static and covalent crystal fields. Projecting the five
lowest eigenstates onto the Ti$^{3+}$ sector (which corresponds to
the states $\big|d^1\big>$), gives to a very good approximation
the same eigenstates as for the static crystal field alone, as can
be seen by comparing  Table \ref{covcf} with Table \ref{tab2}.
This finding explains why, despite the admixture of Ti$^{2+}$
states $\big|d^2\big>$, the agreement with the experiments of
Refs.~\cite{itoh} and \cite{aki} remains perfect. Indeed, these
experiments measure the Ti$^{3+}$ part, $\big|d^1\big>$, of the
combined static and covalent crystal field, and apparently are not
sensitive to the Ti$^{2+}$ admixture $\big|d^2\big>$. Table
\ref{covcf} also shows that the $t_{2g}$ splitting remains almost
the same as in the absence of the covalent contribution, whereas
the distance between the $t_{2g}$ and $e_g$ energies is enhanced.

\begin{table}
\caption{The combined static and covalent crystal field for
Ti$^{3+}$ (site 1): Spectrum and eigenstates in the orthorhombic
basis for the $d$ basis, see Eq.~(\ref{dbasis}).} \label{covcf}
\begin{ruledtabular}
\begin{tabular}{r|rrrrr}
 --0.673 eV & (\hspace*{0.5em}0.187, & --0.340,& --0.438,& 0.583, & --0.564)\\ \hline
--0.519 eV & (--0.028, & --0.350, & 0.573, & 0.622,& 0.402) \\
\hline --0.409 eV & (\hspace*{0.5em}0.459, & 0.274, & 0.634, &
--0.050, & --0.557) \\ \hline
 0.737 eV & (\hspace*{0.5em}0.751, & 0.342, & --0.280, & 0.188, & 0.453) \\ \hline
 0.865 eV & (--0.435, & 0.755, & --0.036, & 0.485, & --0.072)
\end{tabular}
\end{ruledtabular}
\end{table}

We now  discuss the crystal-field gap and the $t_{2g}$ splitting
scheme as  obtained from our calculation and from an alternative
calculation \cite{rueckamp}, in relation with an analysis of the
optical conductivity \cite{rueckamp} and Raman data
\cite{ulrich2}. Our static crystal-field calculation  yields a
non-degenerate orbital ground state separated by $\approx $0.15 eV
from the first excited state and a second excited state separated
by $\approx $0.13 eV  from the first excited one (see Table
\ref{tab2}). This $t_{2g}$ splitting scheme results from the
orthorhombic distortion of the crystal and from the distortion of
the oxygen octahedra. We have estimated that the covalent crystal
field reduces the gap between the first and the second excited
states (to about 0.11 eV according to Table \ref{covcf}), while
the gap between the ground state and the first excited state
remains practically the same. A more precise calculation of the
covalent crystal field \cite{rueckamp}, which takes into account
two additional effects, the $pp$ hybridization and the Ti$^{1+}$
admixture \cite{havpriv}, gives $\approx $0.19 eV for the gap
between the ground state and the first excited state and  $\approx
$0.14 eV for the gap between the first and the second excited
states (the Ti$^{1+}$ admixture means that there is also an
admixture of $d^3$ states to the ground state). This result is in
better agreement with the data of optical conductivity
\cite{rueckamp} and Raman spectroscopy \cite{ulrich2}, which show
that the first orbital excitation is centered around 0.2--0.25 eV.

Since it is extremely complicated to include in the magnetic
superexchange calculation the hopping between the $pd$ hybridized
states, our calculations below contain only the hopping between
the Ti$^{3+}$ states. The results listed in Table \ref{covcf},
which show that the projections of the eigenstates of the combined
static and covalent crystal fields onto the Ti$^{3+}$ sector are
almost the same as in the static-only case, ensure that the
Ti$^{3+}$ ground states we use are an appropriate starting point
for the superexchange calculation.

\subsection{The magnetic moment}

The calculation of the magnetic structure detailed below yields
the directions of the magnetic moments in the ground state, but
does not determine the magnitude of the moment. However, one can
estimate that magnitude by diagonalizing together
$H_{mn}^{\text{cf}}$ and $H_{mn}^{\text{so}}$ for a single
Ti$^{3+}$ ion. The eigenstates of this combined Hamiltonian are
symmetric or antisymmetric with respect to time-reversal, leading
to five Kramers doublets for the single Ti$^{3+}$ ion. We use
those doublets to find the expectation values of the  angular
momentum.  By choosing the largest possible polarization of the
magnetic moment along the $z$ axis (that direction is the leading
one of the observed moment \cite{garret}) out of all the linear
combinations of the ground-state doublet, we find
$\big<l^z_k+2s^z_k\big>\,\mu_{\text{B}}^{\nil}=0.91\,
\mu_{\text{B}}^{\nil}$. This  partially explains the reduction of
the observed ordered moment as compared  to $1\,\mu_{\text{B}}$.

The Ti--O hybridization hardly affects the magnetic moment. For
the parameters used here, the admixture of spin 0 and spin 1
Ti$^{2+}$ states into the ground state of the covalent crystal
field reduces the ordered moment by only $\approx$ 0.1 \%.

\section{The superexchange couplings} \label{secouplings}

Our aim is to obtain from the full Hamiltonian,
$H_{mn}^{\nil}=H_{mn}^0+V_{mn}^{\nil}$, an effective spin
Hamiltonian, $h_{mn}^{\nil}$, which acts within the Hilbert space
of the four-fold degenerate ground state of the unperturbed
Hamiltonian $H_{mn}^0$.

Since the Hamiltonian is invariant under time-reversal,   there
are no single-ion terms, and consequently the effective spin
Hamiltonian, to second-order in the spin variables,  takes the
form
\begin{equation}
h_{mn}={\mathbf{S}}_m \cdot A_{mn} \cdot {\mathbf{S}}_n,
\end{equation}
where $A_{mn}^{\hspace*{0em}}\big( =A_{nm}^t\big)$ is the $3
\times 3$ superexchange matrix.   This matrix may be decomposed
into a symmetric part and an antisymmetric one. The three
components of the latter constitute the Moriya vector
${\mathbf{D}}_{mn}( =-{\mathbf{D}}_{nm}\big)$. Extracting further
the isotropic part of $A_{mn}$, i.\,e., the Heisenberg coupling
$J_{mn}$, the effective spin Hamiltonian is cast into the form
\begin{equation}
h_{mn}^{\hspace*{0em}}=J_{mn}^{\hspace*{0em}}{\mathbf{S}}_m^{\hspace*{0em}}
\cdot
{\mathbf{S}}_n^{\hspace*{0em}}+{\mathbf{D}}_{mn}^{\hspace*{0em}}\cdot
\big({\mathbf{S}}_m^{\hspace*{0em}}\times{\mathbf{S}}_n^{\hspace*{0em}}\big)+
 {\mathbf{S}}_m^{\hspace*{0em}} \cdot A_{mn}^{\text{s}}\cdot
 {\mathbf{S}}_n^{\hspace*{0em}} .\label{magnetich}
\end{equation}
Here, $A_{mn}^{\text{s}}$ represents the symmetric anisotropy. Due
to the space-group symmetries, all three types of magnetic
couplings belonging to the eight planar Ti--Ti bonds may be
obtained from those of a single bond, and so is the case for the
four inter-planar bonds, see Ref.~\cite{us}.

The various magnetic couplings appearing in Eq.~(\ref{magnetich})
are obtained by perturbation theory to leading order in $V_{mn}$,
namely, to  second order in the hopping $t_{mn}$ and to first and
second order in  the spin-orbit coupling (scaled by $\lambda$).
The formal expressions of the perturbation expansion are
documented in  Ref. \cite{us}.  The Heisenberg isotropic exchange
[the first term in Eq.~(\ref{magnetich})] is  independent of
$\lambda$. A systematic description of the magnetic anisotropies
due to the spin-orbit interaction requires both the first and the
second order processes in $\lambda$ \cite{shekht}. The technical
reason being that the expectation value of the cross product in
the second term of Eq.~(\ref{magnetich}) is, in fact, also of
order $\lambda$, so that altogether the Dzyaloshinskii-Moriya
interaction contributes to the exchange energies in at least
second order in the spin-orbit coupling. We neglect terms in which
there appear two Ti$^{2+}$ intermediate states in  the
perturbation expansions. These are smaller than the ones we keep,
by an additional factor of $\simeq
\Delta_{\text{cf}}/U_{\text{eff}}=0.043$,  where
$\Delta_{\text{cf}}=0.150$\,eV is the gap between the ground state
of the single-particle crystal field and the first excited state,
see Table \ref{tab2}. The detailed calculation of the various
terms  is lengthy, albeit straight-forward. More details are given
in Ref.~\cite{us}. The values we obtain, using the parameters
cited above, are listed in Table \ref{microscres}.

\begin{table}
\caption{The calculated single-bond spin couplings (in meV). The
Moriya vectors are given including the  corrections
$\mathbf{D}'_{mn}$, which are of order $\lambda^2$. The symmetric
anisotropies are given as
${\mathbf{A}}_{mn}^{\text{d}}=(A_{mn}^{xx},A_{mn}^{yy},A_{mn}^{zz})$
and
${\mathbf{A}}_{mn}^{\text{od}}=(A_{mn}^{yz},A_{mn}^{xz},A_{mn}^{xy})$
for the diagonal and off-diagonal entries, respectively.}
\label{microscres}
\begin{ruledtabular}
\begin{tabular}{c}
Heisenberg couplings \\ \hline $J_{12}=-3.870,\,J_{13}=2.772$
 \\ \hline Moriya vectors \\ \hline
${\mathbf{D}}_{12}=(1.776, -0.938, -0.325),
\,{\mathbf{D}}_{13}=(-0.671, 0.189 ,0)$
\\ \hline Symmetric anisotropies \\ \hline
${\mathbf{A}}_{12}^{\text{d}}=(0.175,\!-0.011,\!-0.160),
{\mathbf{A}}_{13}^{\text{d}}=(-0.145,\!-0.024,0.010)$,\\
${\mathbf{A}}_{12}^{\text{od}}=(0.044,-0.131,-0.313),
\,{\mathbf{A}}_{13}^{\text{od}}=(0,0,-0.153)$
\end{tabular}
\end{ruledtabular}
\end{table}

As is seen from Table \ref{microscres}, the value we find for the
in-plane Heisenberg coupling, $J_{12}$, roughly agrees with the
experimental one, $\approx$--3 meV \cite{ulrich}. However, the
calculated inter-plane coupling, $J_{13}$, is {\em positive}, in
contradiction with experiment \cite{ulrich} (which yields for that
coupling  $\approx$ --3 meV, too). Generally speaking, there are
antiferromagnetic and ferromagnetic contributions to the isotropic
Heisenberg couplings. The former arise when the intermediate
Ti$^{2+}$ states of the perturbation expansion are singlets, and
the latter when they are triplets. Separating these  competing
contributions, we find $J_{12}^{\rm s}$=21.481 meV, $J_{12}^{\rm
t}$=--25.351 meV, $J_{13}^{\rm s}$=12.008 meV, and $J_{13}^{\rm
t}$=--9.237 meV, namely, the antiferromagnetic and the
ferromagnetic contributions are roughly the same. It is worth
noting that in the case of LaTiO$_3$ \cite{us}, the contribution
of the singlets dominated the one of the triplets, and indeed our
calculation of that compound  has yielded reliable values.
Unfortunately, in the case of YTiO$_3$ the balance between these
competing contributions  is too sensitive to be resolved by our
model  approximations. This delicate balance in the case of
YTiO$_3$ is also reflected in the overall rather small value of
the total isotropic coupling,  $\approx$--3 meV, (whereas it is
15.5 meV in the case of LaTiO$_3$). This value is also very
sensitive to the parameters used. For example, taking
$U_{\text{eff}} \approx 1.6$ eV would have changed the sign of
$J_{13}$. In contrast, a sign change in the case of LaTiO$_3$
requires the considerably lower value of $U_{\text{eff}} \approx
0.6$ eV. We discuss in Sec.~\ref{summary} various difficulties
encountered in obtaining realistic values for the Heisenberg
couplings of YTiO$_3$ which have been previously reported in the
literature.

Had we used the values listed in Table \ref{microscres}, we would
have obtained a predominant A-type antiferromagnetic order for
YTiO$_3$, which  sharply contradicts the experiment \cite{ulrich}.
However, the fact that our values for the leading (isotropic)
superexchange couplings do not agree with experiment does not
necessarily mean that the anisotropic ones are not reliable. In
the case of the cuprates, for example, it has been found (by
comparing with exact diagonalizations)  that while the isotropic
couplings calculated by perturbation theory were inaccurate, the
anisotropic ones were accurate enough \cite{yild}. Since the
latter determine the directions of the spins in the classical
ground state and the spin-wave gaps, a way to test our anisotropic
superexchange couplings is to examine those properties. In order
to do so, we replace in the following the isotropic couplings by
the experimentally-deduced ones,  $J_{12}=J_{13}=-2.75$ meV
\cite{ulrich}, while using for the anisotropic couplings the
values given in Table \ref{microscres}.

\section{The magnetic structure} \label{magngs}

The single-bond spin Hamiltonian, Eq.~(\ref{magnetich}), is the
basis for the magnetic Hamiltonian, from which the magnetic order
of the classical ground state follows. To construct the latter,
the entire Ti-lattice is decomposed into four sublattices. The
four sublattices are enumerated according to the numbers of the
four Ti ions per unit cell shown in Fig.~\ref{fig1} (sublattice
$i=1$ corresponds to Ti ion 1 and its Bravais translations, etc.).
Assigning a fixed magnetization (per site) to all the spins within
each sublattice, ${\mathbf{M}}_i$, one sums over all bonds which
couple the four sublattices, to obtain the {\em macroscopic}
magnetic Hamiltonian in the form
\begin{equation}
H_{\text{M}}^{\nil}=\sum_{ij}\big[I_{ij}^{\hspace*{0em}}
{\mathbf{M}}_i^{\hspace*{0em}}\cdot{\mathbf{M}}_j^{\hspace*{0em}}
+{\mathbf{D}}_{ij}^{\text{D}}\cdot\big({\mathbf{M}}_i^{\hspace*{0em}}
\times{\mathbf{M}}_j^{\hspace*{0em}}\big)+
{\mathbf{M}}_i^{\hspace*{0em}}\cdot
 \Gamma_{ij}^{\nil} \cdot{\mathbf{M}}_j^{\hspace*{0em}}\big],\label{HM}
\end{equation}
where $ij$ runs over the sublattice pairs $12,13,24,$ and $34$ of
Fig.~\ref{fig1}. This summation procedure gives rise to the
macroscopic magnetic couplings: $I_{ij}^{\hspace*{0em}}$ is the
macroscopic isotropic coupling, ${\mathbf{D}}_{ij}^{\text{D}}$ are
the Dzyaloshinskii vectors  which are the macroscopic
antisymmetric anisotropies, and $\Gamma_{ij}^{\nil}$ are the
macroscopic symmetric anisotropy tensors. The relations between
those macroscopic couplings and the microscopic single-bond
couplings and  the inter-relations among the macroscopic
couplings,  dictated by the symmetries of the space-group, are
discussed in  Ref.~\cite{us}.

We now  minimize $H_{\text{M}}$, and find the various sublattice
magnetizations. We assume that all four vectors ${\mathbf{M}}_i$
have the same magnitude,  $M$, but differ in their directions.
Since Eq.~(\ref{HM}) is quadratic in $M$, the minimization will
only yield the directions of these vectors, and not the value of
$M$. To simplify the procedure we use group theory. According to
the space group $Pbnm$ symmetries, there are four possibilities
for the symmetry of sublattice magnetizations of the classical
ground state, as listed in Ref.~\cite{meijer}. Having checked each
of them, we have concluded that only one of these possibilities
has the lowest energy.  We then find that  the classical magnetic
ground state has the following structure: The $x$ components of
the magnetizations order antiferromagnetically, in a G-type
structure (where the four sublattices actually reduce to two). The
$y$ components order antiferromagnetically as well, but in an
A-type structure. Finally, the $z$ components of the
magnetizations order ferromagnetically. This structure agrees with
experiment \cite{ulrich}. This is a non-trivial result caused by
the anisotropic spin couplings. Given only the ferromagnetic
Heisenberg couplings, the ferromagnetic moment could also be
oriented along the $x$ or the $y$ axis, see Ref.~\cite{meijer}.

\begin{table}
\caption{ The structure of the magnetic order in
YTiO$_{\text{3}}$, characterized in terms of the sublattice
magnetizations ${\mathbf{M}}_i$ in the classical ground state
(normalized to the value $M$), which are expressed by the canting
angles $\varphi$ and $\vartheta$.  } \label{cgs}
\begin{ruledtabular}
\begin{tabular}{c}
 $x$ components: G-type\\ \hline
$-M_1^x=M_2^x=M_3^x=-M_4^x=M\cos \varphi \sin \vartheta $\\
\hline $y$ components: A-type\\ \hline
$-M_1^y=-M_2^y=M_3^y=M_4^y=M\sin \varphi \sin \vartheta $\\
\hline $z$ components: ferromagnetic \\ \hline
$M_1^z=M_2^z=M_3^z=M_4^z=M\cos \vartheta $
\end{tabular}
\end{ruledtabular}
\end{table}

The magnetic structure can be specified by expressing the four
magnetizations in terms of two canting angles, $\varphi$ and
$\vartheta$, see Tables \ref{cgs} and \ref{macroscres}. The angle
$\vartheta$, measured with respect to the $z$ axis, is
proportional to the spin-orbit parameter $\lambda$ (as found by
varying this parameter), while the angle $\varphi$ is almost {\em
independent} of it. Indeed, one may verify that for an
infinitesimally small $\lambda$, there is just a ferromagnetic
order along the $z$ axis. As $\lambda$  increases, so does
$\vartheta$, causing an increasing canting of the magnetizations.
However, the projection of the magnetic moment onto the $xy$
planes remains almost perpendicular to the rotation axis of the
magnetization, and hence $\varphi$ is practically unaffected by
the value of $\lambda$. Interestingly enough, the magnetic
structure of LaTiO$_3$ can also be described in terms of such
canting angles. However, in that case both $\varphi$ and
$\vartheta$ (the latter measured with respect to the $xy$ planes)
are proportional to $\lambda$ \cite{us}, leading to a (mainly)
G-type order along the $x$ direction which would have occurred
even for an infinitesimally small $\lambda$.

It is worth noting that using naively the procedure outlined above
to obtain the energy of the classical magnetic ground-state might
yield non-systematic contributions up to fourth order in the
spin-orbit coupling $\lambda$ \cite{us,shekht}. To exemplify this
point, we consider the expectation value of $H_{\text{M}}$,
expressed in terms of the canting angles $\varphi$ and
$\vartheta$, \setlength{\arraycolsep}{0cm}
\begin{widetext}
\begin{eqnarray}
\hspace*{-0.8cm} \big<H_{\text{M}}^{\nil}\big>= \big[\lambda^0\!
:\big]&& \;\;\,\,2
(I_{12}^{\nil}+ I_{13}^{\nil})\cos^2 \vartheta \nonumber\\
\big[\lambda^2\!:\big]&& -\,2(I_{12}^{\nil}+I_{13}^{\nil})\cos^2
\varphi \sin^2 \vartheta + 2(I_{12}^{\nil}-I_{13}^{\nil})\sin^2
\varphi \sin^2 \vartheta  \nonumber \\
&& +\,4(D_{12}^{\text{D}\,y}+ D_{13}^{\text{D}\,y}) \cos \varphi
\cos \vartheta \sin \vartheta -4D_{13}^{\text{D}\,x}\sin \varphi
\cos \vartheta  \sin \vartheta
+2 (\Gamma_{12}^{zz}+\Gamma_{13}^{zz})\cos^2 \vartheta \nonumber\\
\big[\lambda^3\!:\big]&& +\,4D_{12}^{\text{D}\,z}\cos \varphi \sin
\varphi \sin^2 \vartheta  -2 \Gamma_{12}^{yz} \sin
\varphi \cos \vartheta  \sin \vartheta \nonumber \\
\big[\lambda^4\!:\big]&&  -\,2(\Gamma_{12}^{xx}+\Gamma_{13}^{xx})
\cos^2 \varphi \sin^2 \vartheta
+2(\Gamma_{12}^{yy}-\Gamma_{13}^{yy}) \sin^2 \varphi \sin^2
\vartheta -4\Gamma_{13}^{xy} \cos \varphi  \sin \varphi \sin^2
\vartheta . \nonumber \\ \label{clgsenY}
\end{eqnarray}
\end{widetext}
We take the contributions up to the order $\lambda^3$ into
account, i.\,e., we {\em{exclude}}  from the calculation of the
classical ground state the coefficients  $\Gamma_{12}^{xx}$,
$\Gamma_{13}^{xx}$, $\Gamma_{12}^{yy}$, $\Gamma_{13}^{yy}$,
$\Gamma_{13}^{xy}$, and the $\lambda^2$ correction of
$D_{12}^{\text{D}\,z}$. This procedure yields the
{\em{macroscopic}} magnetic couplings listed in Table
\ref{macroscres}. Using these couplings we have calculated the
canting angles $\varphi$ and $\vartheta$, and the ordered magnetic
moments. These results are also listed in Table \ref{macroscres}.

\begin{table}
\caption{The macroscopic magnetic couplings in meV, the resulting
canting angles of the magnetizations in the classical ground
state, and the resulting values of the ordered moments (normalized
to the value $M$). Three coefficients of the macroscopic symmetric
anisotropies are taken into account (see text).}
\label{macroscres}
\begin{ruledtabular}
\begin{tabular}{c}
Isotropic couplings \\ \hline $I_{12}=-2.750,\,I_{13}=-1.375$ \\
\hline Dzyaloshinskii vectors \\ \hline
${\mathbf{D}}_{12}^{\text{D}}=(0,-0.938 ,-0.367 ),
\,{\mathbf{D}}_{13}^{\text{D}}=(-0.335, 0.094,0)$  \\ \hline
Macroscopic symmetric anisotropies \\
\hline $\Gamma_{12}^{zz}=-0.160,
\,\Gamma_{13}^{zz}=0.005,\,\Gamma_{12}^{yz}=0.044$ \\ \hline
Canting angles
\\ \hline
 $\varphi=-44.17^\circ,\,\vartheta=7.55^\circ$ \\ \hline
Ordered moments \\ \hline ${\mathbf{M}}=(\pm 0.094, \pm
0.092,0.991)\,M$
\end{tabular}
\end{ruledtabular}
\end{table}

In order to compare our magnetic structure with the one found
experimentally, we normalize the moments to 1 $\mu_{\text{B}}$.
Then, according to Ref.~\cite{ulrich}, experiment gives
${\mathbf{M}}=(\pm 0.149,\pm 0.085,0.985)\,\mu_{\text{B}}$, with
relative errors of 15 \% for the G-type moment, 25 \% for the
A-type moment, and 2 \% for the ferromagnetic moment. The
calculated values are within a single error bar except for the
G-type moment for which we obtain a value which is 37 \% lower
than the experimental one (i.e., within the $3 \sigma$ range of
the measurement). Thus, the calculated magnetic structure is in
reasonable agreement with experiment.

\section{The spin-wave spectrum} \label{spinwaves}

\subsection{The spin-wave Hamiltonian}

The calculation of the spin-wave dispersion is carried out
analogously to the case of LaTiO$_3$ \cite{us}, since all
symmetries  are the same for both YTiO$_3$ and LaTiO$_3$. As in
the calculation of the classical magnetic ground state,  we
combine the experimental Heisenberg couplings
$J_{12}=J_{13}=-2.75$ meV \cite{ulrich} with our calculated
anisotropic couplings which contribute systematically to the
classical ground-state energy [see Eq.~(\ref{clgsenY}) and the
following discussion].

Since the classical magnetic ground state is characterized by four
sublattices, we will obtain four branches in the spin-wave
dispersion. The first step in the standard calculation of
spin-wave dispersions is the rotation of the local coordinates at
each sublattice, $i$, such that the new $z$ axis will point in the
direction of the corresponding sublattice ground-state
magnetization, ${\mathbf{M}}_i$. This rotation still leaves  the
freedom to choose the new local $x$ and $y$ axes, i.e., to rotate
the new coordinate system around its $z$ axis. Denoting the new
local coordinate system by $x'_{i}$, $y'_{i}$ and $z'_{i}$
($i=1,2,3,4$), we find that the convenient choice for our purpose
is
\begin{eqnarray}
\hat{z}'_{i}&=&\frac{{\mathbf{M}}_i}{M}, \ \
\hat{y}'_{i}=\frac{{\mathbf{M}}_{i}\times\hat{x}}{m_{i}}, \ \
\hat{x}'_{i}=\hat{y}'_{i}\times\hat{z}'_{i}, \ \
M=\big|{\mathbf{M}}_i\big|, \nonumber \\
m_{i}&=&\sqrt{(M_{i}^{y})^{2}+(M_{i}^{z})^{2}}.\label{localcoor}
\end{eqnarray}
In the rotated coordinate system the spin Hamiltonian, comprising
all three types of magnetic couplings, takes the form
\begin{eqnarray}
h=\sum_{\langle mn\rangle}{\mathbf{S}}_m' \cdot  A_{mn}' \cdot
{\mathbf{S}}_n', \label{trafo}
\end{eqnarray}
where the primes denote  the rotated quantities. In particular,
$A'_{mn}$ is the 3$\times$3 superexchange matrix in rotated
coordinates.

Since we consider only the Ti ions, it is convenient to use a
coordinate system in which the Ti ions occupy the sites of a
simple cubic lattice, of unit lattice constant (this picture is
the appropriate one for comparing with the experimental spin-wave
data \cite{ulrich}, as discussed in the next subsection). It is
also convenient to use a coordinate system in which
nearest-neighbor Ti ions are located along the axes (namely, to
rotate the orthorhombic coordinates by $-45^\circ$ around the $z$
axis, see Fig.~\ref{fig1}). Accordingly, our magnetic unit cell is
spanned by the vectors $(1,1,0)$, $(1,-1,0)$, and $(0,0,2)$,  and
the corresponding magnetic Brillouin zone (MBZ) is defined by
\begin{widetext}
\begin{eqnarray}
|q_{x}+q_{y}|\leq \pi, \ \ \ |q_{z}|\leq \frac{\pi}{2}.\label{BZ}
\end{eqnarray}

The resulting spin-wave dispersion (for more details of the
derivation, see Ref.~\cite{us}) consists of four branches,
\begin{eqnarray}
\Omega _{1}^{2}({\bf q})&=&(C_{1}+C_{2}^{\perp}\cos q_{z})^{2}
-|C_{3}^{\perp}|^{2}\cos^{2}q_{z}+|C_{2}^{\parallel}|^{2}(\cos
q_{x}+\cos q_{y})^{2}
-|C^{\parallel}_{3}\cos q_{x}+C^{\parallel *}_{3}\cos
q_{y}|^{2}\nonumber \\ &-&(\cos q_{x}+\cos q_{y})W(-\cos q_{z}),\nonumber\\
\Omega_{2}^{2}({\bf q})&=&\Omega_{1}^{2}({\bf q}+{\bf Q}),\ \ \
{\rm with}\ \ {\bf Q}=(0,0,\pi ),\nonumber\\
\Omega_{3}^{2}({\bf q})&=&\Omega_{1}^{2}({\bf q}+{\bf Q}'),\ \ \
{\rm with}\ \ {\bf Q}'=(\pi,\pi,0 ),\nonumber\\
\Omega_{4}^{2}({\bf q})&=&\Omega_{1}^{2}({\bf q}+{\bf Q}''),\ \ \
{\rm with}\ \ {\bf Q}''={\bf Q}+{\bf Q}'=(\pi,\pi,\pi
),\label{omegas}
\end{eqnarray}
where
\begin{eqnarray}
W^{2}(\cos q_{z})&=&4\Bigl [(C_{1}-C_{2}^{\perp}\cos
q_{z})^{2}-|C^{\perp}_{3}|^{2}\cos^{2}q_{z}\Bigr ]\Bigl
[|C_{2}^{\parallel}|^{2}-\Bigl
(\frac{C_{3}^{\parallel}+C_{3}^{\parallel *}}{2}\Bigr )^{2}\Bigr
]\nonumber\\
&+&\Bigl [(C^{\perp
*}_{3}C^{\parallel}_{2}+C^{\perp}_{3}C^{\parallel *}_{2})\cos
q_{z}+(C_{1}-C_{2}^{\perp}\cos
q_{z})(C^{\parallel}_{3}+C^{\parallel *}_{3})\Bigr ]^{2}.
\label{WWW}
\end{eqnarray}
Each of the spin-wave branches has tetragonal symmetry, i.e.,
$\Omega_i(q_x,q_y,q_z)=\Omega_i(q_y,q_x,q_z)$\\ $=
\Omega_i(-q_x,q_y,q_z)=\Omega_i(q_x,-q_y,q_z)=\Omega_i(q_x,q_y,-q_z)
$. The coefficients in Eqs. (\ref{omegas}) and (\ref{WWW}) are
linear combinations of the coefficients $C_{mn}(\ell)$,
\begin{eqnarray}
C_1^{\nil}&=&2C_{13}^{\nil}(1)+4C_{12}^{\nil}(1)=C_1^{*}, \quad
C_2^{\perp }=2C_{13}^{\nil}(2)=C_2^{\perp *},\quad
C_2^{\|}=2C_{12}^{\nil}(2), \quad C_3^{\perp }=2C_{13}^{\nil}(3),
\quad C_3^{\| }=2C_{12}^{\nil}(3). \label{swcoeff}
\end{eqnarray}
These are given by combinations of the superexchange matrix
elements $(A'_{mn})^{\alpha\beta}$,
\begin{eqnarray}
C_{mn}^{\nil}(1)&=&-\,\mbox{$\frac{1}{2}$}(A_{mn}')^{zz}, \nonumber  \\
C_{mn}^{\nil}(2)&=&\quad \mbox{$\frac{1}{4}$}\big[(A_{mn}')^{xx}+
(A_{mn}')^{yy}+i\big((A_{mn}')^{yx}-(A_{mn}')^{xy}\big)\big], \nonumber  \\
C_{mn}^{\nil}(3)&=&\quad \mbox{$\frac{1}{4}$}\big[(A_{mn}')^{xx}-
(A_{mn}')^{yy}+i\big((A_{mn}')^{yx}+(A_{mn}')^{xy}\big)\big].
\label{BosSBHam1}
\end{eqnarray}
\end{widetext}
The explicit expressions are  not reproduced here since their
expressions are very long.

Equations (\ref{omegas}) contain our final result for the
spin-wave spectrum of YTiO$_{3}$. Evidently, the details of the
spectrum can be obtained only numerically: One has to write the
spin-wave coefficients, Eqs.~(\ref{swcoeff}), in terms of those
appearing in Eqs.~(\ref{BosSBHam1}),  and express the latter in
terms of the original coefficients of the spin Hamiltonian
(\ref{magnetich}). These results are then used in constructing the
dispersion.

When the spin-orbit parameter $\lambda$ is set to zero the
coefficients appearing in Eqs.~(\ref{omegas}) simplify to
\begin{eqnarray}
C_{1}^{\nil}&=&-2J_{12}^{\nil}-J_{13}^{\nil}, \,
C^{\perp}_{2}=J_{13}^{\nil}, \, C^{\parallel}_{2}=J_{12}^{\nil},
\, C^{\perp}_{3}=C^{\parallel}_{3}=0, \nonumber \\
\end{eqnarray}
where $J_{12}<0$ is the in-plane Heisenberg coupling, and
$J_{13}<0$ is the Heisenberg coupling between planes. In that case
\begin{eqnarray}
\Omega_{1}^{2}({\bf q})&=&\big[2J_{12}^{\nil}+J_{13}^{\nil}-J_{12}^{\nil}(\cos q_x^{\nil}+\cos q_y^{\nil})-J_{13}^{\nil}\cos q_z^{\nil}\big]^2, \nonumber \\
\Omega_{2}^{2}({\bf q})&=&\big[2J_{12}^{\nil}+J_{13}^{\nil}-J_{12}^{\nil}(\cos q_x^{\nil}+\cos q_y^{\nil})+J_{13}^{\nil}\cos q_z^{\nil}\big]^2, \nonumber \\
\Omega_{3}^{2}({\bf q})&=&\big[2J_{12}^{\nil}+J_{13}^{\nil}+J_{12}^{\nil}(\cos q_x^{\nil}+\cos q_y^{\nil})-J_{13}^{\nil}\cos q_z^{\nil}\big]^2, \nonumber \\
\Omega_{4}^{2}({\bf
q})&=&\big[2J_{12}^{\nil}+J_{13}^{\nil}+J_{12}^{\nil}(\cos
q_x^{\nil}+\cos q_y^{\nil})+J_{13}^{\nil}\cos q_z^{\nil}\big]^2.
\nonumber \\
\end{eqnarray}
Only $\Omega_{1}({\bf q})$ vanishes at the zone center and is
hence termed the acoustic mode. The other branches have gaps at
the zone center: $\Omega_{2}({\bf 0})=2|J_{13}|$, $\Omega_{3}({\bf
0})=4|J_{12}|$, $\Omega_{4}({\bf 0})=4|J_{12}|+2|J_{13}|$, and are
hence   termed the optical modes. Indeed, when only the
ferromagnetic couplings $J_{12}$ and $J_{13}$ are kept (i.e., for
$\lambda =0$),  the magnetic unit cell contains only one Ti ion,
corresponding a simple cubic lattice. The Brillouin zone  is then
four times as large as the Brillouin zone of Eq.~(\ref{BZ}). By
''folding out'' the three optical modes into the larger Brillouin
zone, one reproduces the usual gapless dispersion of the pure
(ferromagnetic) Heisenberg model. At finite values of the
spin-orbit coupling all modes have gaps at the zone center, but
the one of $\Omega_{1}$ is much smaller than those of the other
three modes.

\subsection{Numerical results for the spin-wave dispersion}

\subsubsection{The acoustic branch}

\begin{figure}
(a)\\[1ex]
\includegraphics[width=7cm]{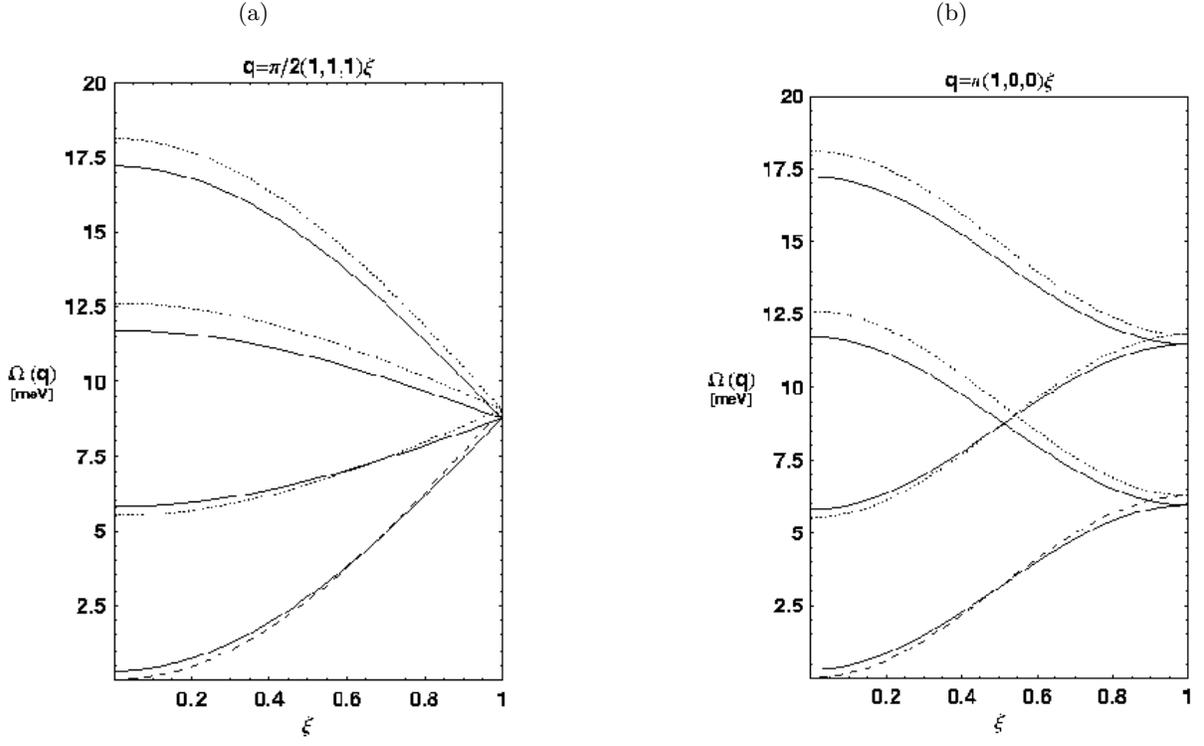}
\caption{The spin-wave dispersion along selected directions in the
magnetic Brillouin zone. Panels (a)--(c) show the four branches
$\Omega_i({\mathbf{q}})$ of our calculated dispersion (solid
curves), the single branch $\Omega({\mathbf{q}})$ (dashed curves)
which has been fitted onto neutron-scattering experiments,
Eq.~(\ref{expfitY}), according to Ref.~\cite{ulrich}, and three
branches, which are obtained from $\Omega({\mathbf{q}})$ by
folding it back from the MBZ of the uncanted ferromagnet into the
smaller MBZ of the canted ferromagnet (dotted curves, see text).
(a) The dispersion along the direction (1,1,1); (b) the dispersion
along (1,0,0); (c) the dispersion along (0,0,1).} \label{fig3}
\end{figure}

Using an isotropic ferromagnetic nearest-neighbor Heisenberg
coupling $J=-2.75\,\text{meV}$ for all bonds,  an anisotropy
parameter $A=0.8\,\text{meV}$ (which expresses the symmetric
anisotropies which are allowed in a cubic situation), and a
zone-center gap $\Delta=0.093 \frac{A^2}{|J|}=0.02\,\text{meV}$,
the authors of Ref.~\cite{ulrich} have fitted the measured
neutron-scattering data numerically onto the dispersion,
\begin{widetext}
\begin{eqnarray}
\Omega({\mathbf{q}}) \simeq &&\sqrt{|J| [3 \!-\! (\cos q_x  \!+\!
\cos q_y \!+\!  \cos q_z)]\!+\!\Delta\!+\!A(1\!-\!\cos q_x)}
\sqrt{|J| [3 \!-\! (\cos q_x \!+\! \cos q_y \!+\!  \cos
q_z)]\!+\!\Delta\!+\!A(1\!-\!\cos q_y)}. \; \label{expfitY}
\end{eqnarray}
\end{widetext}
The numerically-fitted zone-center gap is extremely small. On the
other hand, Ref.~\cite{ulrich} reports an upper bound, 0.3 meV,
for the gap $\Delta$. We find for the acoustic branch
\begin{eqnarray}
\Delta_1=\Omega_1({\mathbf{0}})=0.326\,\text{meV}.
\end{eqnarray}
This value for the zone-center gap roughly agrees with the upper
bound according to Ref.~\cite{ulrich}. A more severe discrepancy
concerns the anisotropy parameter $A$. This parameter implies that
the diagonal and off-diagonal entries of the symmetric anisotropy
tensors are given by
\begin{eqnarray}
{\bf A}_{12}^{\rm
d}&=&\mbox{$\frac{1}{2}$}(A,A,0)=(0.4,0.4,0)\,{\rm meV}, \nonumber
\\
{\bf A}_{13}^{\rm d}&=&(0,0,A)=(0,0,0.8)\,{\rm meV}, \nonumber \\
{\bf A}_{12}^{\rm
od}&=&\mbox{$\frac{1}{2}$}(0,0,-A)=(0,0,-0.4)\,{\rm meV},
\nonumber \\
{\bf A}_{13}^{\rm od}&=&(0,0,0). \label{expsymaniso}
\end{eqnarray}
This result is in contrast with our calculated values for the
symmetric anisotropy tensors according to Table \ref{microscres}.
However, as is noted above [see Eq. (\ref{clgsenY})] and also
elsewhere \cite{us}, {\em both} antisymmetric {\em{as well as}}
symmetric anisotropies contribute to the same order in the
spin-orbit parameter to the classical ground-state energy and
hence to the spin-wave gap and its dispersion. In other words, it
is not possible to express all anisotropies in terms of a single
parameter and the zone-center gap, or alternatively, it is not
possible to deduce the strength of the Dzyaloshinskii-Moriya
interaction directly from the spin-wave dispersion. A proper
procedure is to compare the full measured and calculated
dispersions. Figures \ref{fig3} (a)--(c) show such a comparison.
The agreement of the calculated acoustic branch with the
experimental function of Eq. (\ref{expfitY}) is satisfying (though
it is quantitatively not as good as in case of LaTiO$_3$). The
calculated tetragonal anisotropy of the acoustic branch is found
to be
\begin{eqnarray}
\frac{\Omega_1(0,0,\mbox{$\frac{\pi}{2}$})}
{\Omega_1(\mbox{$\frac{\pi}{2}$},0,0)}=97.74\,\%.
\end{eqnarray}
This value is smaller compared to the one found for LaTiO$_3$
\cite{us}, mainly because in the present case the Heisenberg
couplings are taken to be isotropic over the lattice.

\begin{figure}
(b)\\[1ex] \includegraphics[width=7cm]{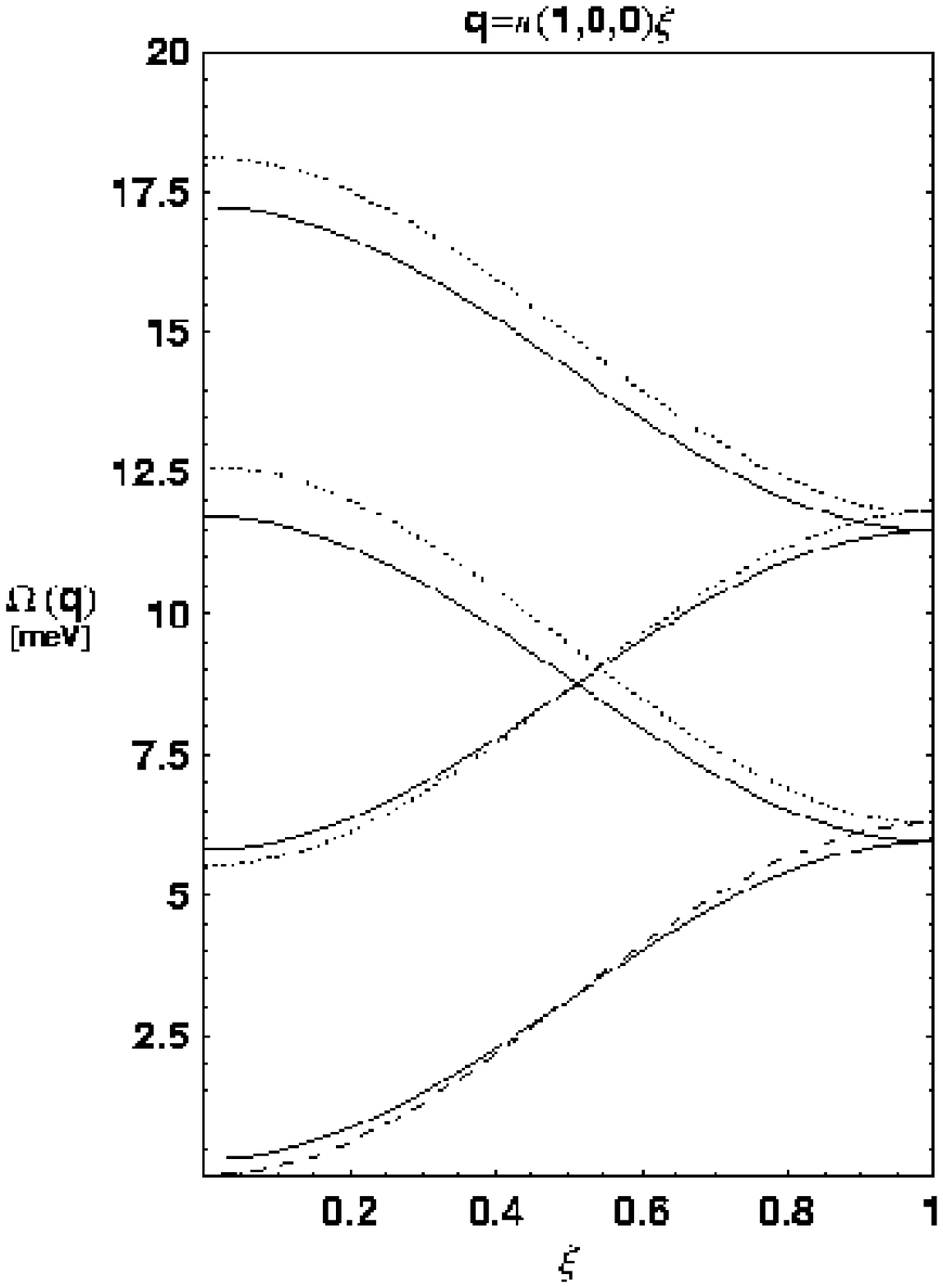}
\end{figure}

\begin{figure}
(c)\\[1ex]
\includegraphics[width=7cm]{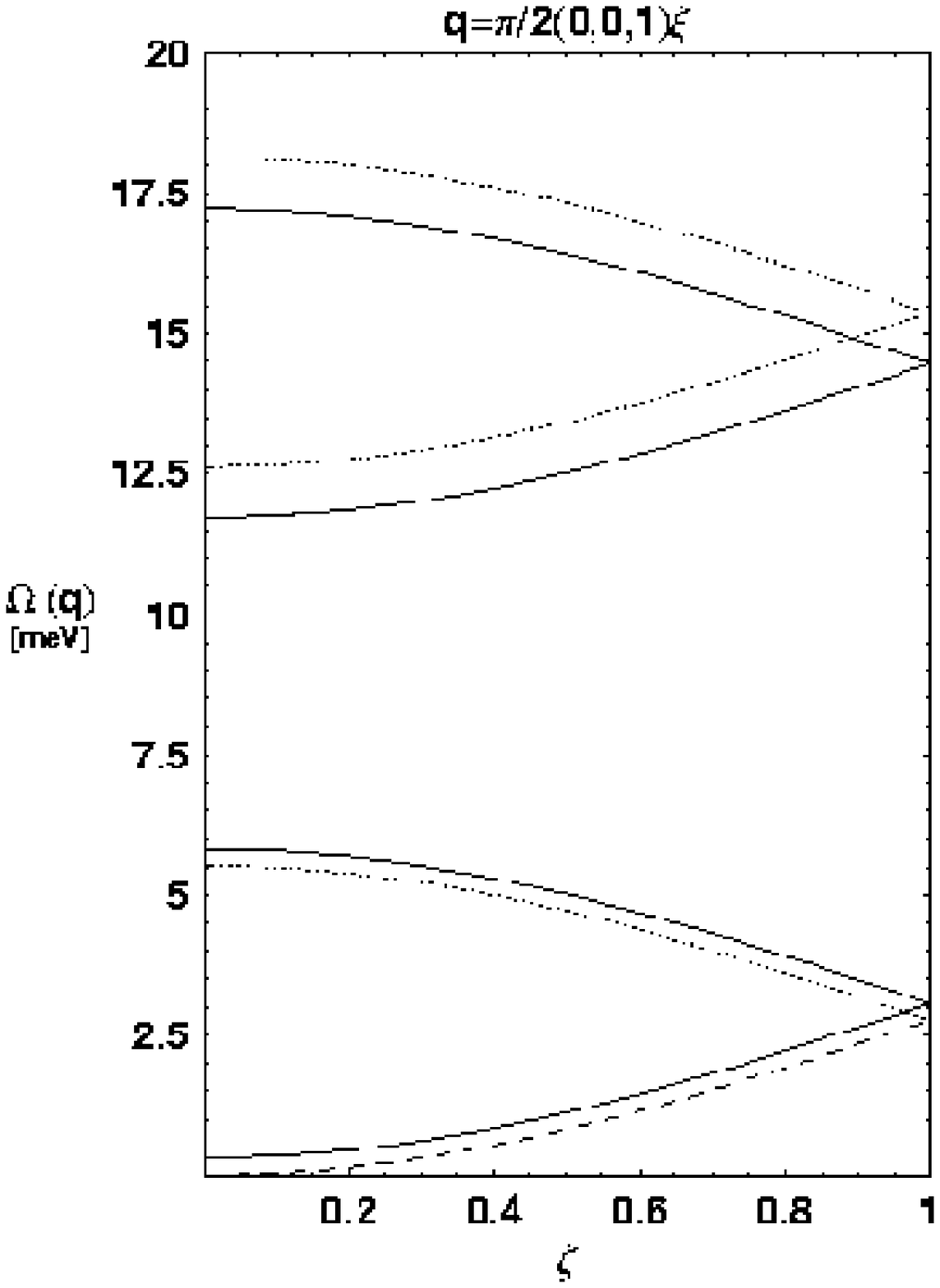}
\end{figure}

\subsubsection{The optical branches}

There is experimental evidence for optical spin-wave branches,
though this has not been pointed out explicitly in
Ref.~\cite{ulrich}. There, a plot of the dispersion along the
(0,0,1) direction (in the pseudocubic coordinates  also used here)
is shown. The plot range includes wave vectors (in our notation)
${\mathbf{q}}=\pi/2(0,0,1)\xi$ with $\xi=0...2$. One may note
however, that for $\xi > 1$ these wave vectors are located
{\em{outside}} the (first) MBZ, see Eq.~(\ref{BZ}). The reason is
that ${\mathbf{Q}}=(0,0,\pi)$ [as well as
${\mathbf{Q'}}=(\pi,\pi,0)$] is a site on the reciprocal lattice
of sublattice No.~1 and thus, is equivalent to zero wave vector.
Were YTiO$_3$  a ferromagnet without any spin canting, then all
four magnetic sublattices would have combined to form a single
lattice, and ${\mathbf{Q}}$, ${\mathbf{Q}'}$, and
${\mathbf{Q}''}={\mathbf{Q}}+{\mathbf{Q}'}$ would not be sites of
the reciprocal lattice. In this hypothetical case the MBZ would be
given by $|q_x|,|q_y|,|q_z|\leq \pi$. In the actual case, with the
spin canting, the MBZ is four times smaller than that. We prefer
to fold back the data from Ref.~\cite{ulrich} to this smaller MBZ,
i.~e., not to consider only the experimental fit function
$\Omega({\mathbf{q}})$,  but also
$\Omega({\mathbf{q}}+{\mathbf{Q}})$,
$\Omega({\mathbf{q}}+{\mathbf{Q}'})$, and
$\Omega({\mathbf{q}}+{\mathbf{Q}''})$. We have plotted
Figs.~\ref{fig3} (a)--(c) using the back-folded MBZ. The
experimental curves which are folded back into the MBZ of the
canted ferromagnet are in a satisfying agreement with our
calculated optical branches. In fact, the signature of the optical
spin-wave mode with the largest zone-center gap can be seen in the
plots of Ref.~\cite{ulrich}. There, it is related to the wave
vector ${\mathbf{Q}''}$ (which is equivalent to zero wave vector)
and has an energy of about 18 meV.

We suggest to re-analyze also the spin-wave data on LaTiO$_3$
\cite{keimer}, i.~e., to fold the experimentally-deduced
dispersion $\Omega({\mathbf{q}})$ from the MBZ of the
antiferromagnet without spin canting back to the MBZ of the canted
antiferromagnet (which is half as large), considering also
$\Omega({\mathbf{q}}+\mathbf{Q})$. Then one obtains the result
that the optical branches which have been calculated in
Ref.~\cite{us} are consistent with the experimental fit function
of Ref.~\cite{keimer} for the dispersion.

In the following, we summarize the properties of the calculated
optical branches of YTiO$_3$. They have considerable zone-center
gaps,
\begin{eqnarray}
\Delta_2&=&\Omega_2({\mathbf{0}})=5.815\,\text{meV}, \;
\Delta_3=\Omega_3({\mathbf{0}})=11.721\,\text{meV}, \nonumber \\
\Delta_4&=&\Omega_4({\mathbf{0}})=17.214\,\text{meV}.
\end{eqnarray}
Two of the calculated optical branches have considerable
tetragonal anisotropies,
\begin{eqnarray}
\frac{\Omega_2(0,0,\mbox{$\frac{\pi}{2}$})}
{\Omega_2(\mbox{$\frac{\pi}{2}$},0,0)}&=&35.48\,\%, \quad
\frac{\Omega_3(0,0,\mbox{$\frac{\pi}{2}$})}
{\Omega_3(\mbox{$\frac{\pi}{2}$},0,0)}=163.29\,\%, \nonumber \\
\frac{\Omega_4(0,0,\mbox{$\frac{\pi}{2}$})}
{\Omega_4(\mbox{$\frac{\pi}{2}$},0,0)}&=&100.77\,\%.
\end{eqnarray}

As is seen in Figs.~\ref{fig3} (a)--(c), all four spin-wave modes
are highly non-degenerate over a wide range of the MBZ. In
contrast, in the case of LaTiO$_3$ we have found \cite{us} that
the four modes constitute two pairs of quasi-degenerate branches.
The reason for this difference between the two systems is related
to the smallness of the angle $\varphi$ in LaTiO$_3$ (as opposed
to its significant value in the case of YTiO$_3$, see Table
\ref{macroscres}). In LaTiO$_3$ there is a nearly full
translational symmetry, leading to the quasi-degeneracy of the
modes.

\section{Summary and discussion} \label{summary}

We have presented a detailed model that aims to describe the
orbital and the magnetic orders in YTiO$_3$. While the orbital
order that we have calculated turns out to agree very well with
experiment, this is not the case for the magnetic superexchange
couplings: To the lowest order in perturbation theory, we find
that the approximate isotropic coupling between the $ab$ planes is
antiferromagnetic, while experiment indicates that it is
ferromagnetic. This discrepancy is apparently due to a strong
competition between ferromagnetic and antiferromagnetic
contributions to that coupling. The approximations we employ are
not sensitive enough to resolve successfully this competition. In
particular, the neglect of exchange processes which involve double
$p$ holes in the intermediate states, and of the Ti$^{2+}$
admixture into the ground state caused by the covalent crystal
field may be detrimental to the calculation of the isotropic
magnetic coupling.

In fact, the titanates are notorious for the difficulties one
encounters when trying to microscopically derive their properties.
For example, Ref.~\cite{mochi2} finds  a predominant A-type
antiferromagnetic coupling for LaTiO$_3$ while Ref. \cite{mochi3}
predicts a ferromagnetic one, both contradicting the
experimentally detected G-type coupling of that material. Our work
on that compound \cite{us} has yielded the correct magnetic order,
but the application of the same model to YTiO$_3$ turns out to be
not so successful. Similar problems have been reported in other
studies of YTiO$_3$. Reference \cite{solovyev}, while deriving
ferromagnetic couplings, predicts  (in contradiction to the
experiment) a strong anisotropy  between  the intra and the
inter-plane couplings,  i.e., $J_{12}=-2.0$ meV and $J_{13}=-0.6$
meV. Reference \cite{ulrich} finds  antiferromagnetic values for
both these couplings in a parameter range which is considered to
be realistic. Both these papers use models which are different
than ours, but they also employ perturbation theory to second
order in the Ti--Ti hopping to derive the required superexchange
parameters.

It should be emphasized, however, that the starting point of our
model, i.e., the crystal field and the orbital ordering it implies
do give a faithful description for YTiO$_3$. The failure of our
model in producing correctly the isotropic Heisenberg coupling
between the $ab$ planes is likely to be related to the use of low
order perturbation theory and to subtle inaccuracies in the
parameters used. The alternative possibility suggested in
Ref.~\cite{ulrich} based on orbital fluctuations is, in our
opinion, not adequate, since it defies the experimentally-detected
orbital order of the ground state.

In view of the above difficulties, and since it is known that
perturbation theory may be insufficient for the leading isotropic
couplings but may well be reliable for the anisotropic ones, we
have combined together the experimentally-deduced isotropic
couplings of YTiO$_3$ with the computed anisotropic ones, to
calculate the classical magnetic ground state. The result turns
out to be satisfactory, when compared with experiment. Similarly
to LaTiO$_3$, we obtain a G-type moment along the crystallographic
$a$ axis, an A-type moment along the $b$ axis, and a ferromagnetic
moment along the $c$ axis, the latter being the predominant one.
Remarkably, this detailed structure is caused by the anisotropies,
and cannot be derived solely on the basis of symmetry arguments.

An even further check of our procedure is provided by the
calculation of the spin-wave excitations. We find four
dispersions: Three of them have considerable zone-center gaps, 6
meV, 12 meV, and 17 meV, while the fourth one has a very small
gap, of the order of 0.3 meV, and  is approximately isotropic over
the magnetic Brillouin zone. We  have demonstrated that all
branches have experimental counterparts as can be deduced from
neutron-scattering data. Comparing the calculated dispersion with
the experimental one, we have found that they are in  a plausible
agreement.

\begin{acknowledgements}
 We gratefully acknowledge discussions with M. Braden,
 M. Gr\"uninger, A.~B. Harris, M.\,W. Haverkort, and A. Komarek. This work was
 partially supported by the German-Israeli Foundation for
 Scientific Research and Development (GIF).
\end{acknowledgements}

\end{document}